\documentclass[aps,prl,preprint,tightenlines,superscriptaddress,showpacs,byrevtex]{revtex4}
\usepackage{graphicx} % Include figure files
\usepackage{dcolumn}  % Align table columns on decimal point
\usepackage{color}
\definecolor{red}{rgb}{1.0,0.0,0.0}

\graphicspath{{ps}}

\begin{document}

%\vspace*{-3\baselineskip}
%\resizebox{!}{3cm}{\includegraphics{belle.eps}}

\preprint{\vbox{ \hbox{   }
                 \hbox{}
                 \hbox{BELLE-CONF-0714}
               %  \hbox{hep-ex nnnn, if available}
}}

\title{ \quad\\[0.5cm]  Precise measurement of hadronic $\tau$-decays in modes with $\eta$ mesons}

%%% Paper:    
%%% Journal:  Summer 2007 Conference Papers
%%% July 17, 2007
%%% Contacts: 
%%% Non-responding authors or those who said NO are commented out.
%%% ====================================================================
%%% Click the RELOAD button on your web browser to see the updated file.
%%% ====================================================================
%%% Use \input{author} to insert this material into your latex file.
%%%%% Force institutions to appear in alphabetical order when typeset.
\affiliation{Budker Institute of Nuclear Physics, Novosibirsk}
\affiliation{Chiba University, Chiba}
\affiliation{University of Cincinnati, Cincinnati, Ohio 45221}
\affiliation{Department of Physics, Fu Jen Catholic University, Taipei}
\affiliation{Justus-Liebig-Universit\"at Gie\ss{}en, Gie\ss{}en}
\affiliation{The Graduate University for Advanced Studies, Hayama}
\affiliation{Gyeongsang National University, Chinju}
\affiliation{Hanyang University, Seoul}
\affiliation{University of Hawaii, Honolulu, Hawaii 96822}
\affiliation{High Energy Accelerator Research Organization (KEK), Tsukuba}
\affiliation{Hiroshima Institute of Technology, Hiroshima}
\affiliation{University of Illinois at Urbana-Champaign, Urbana, Illinois 61801}
\affiliation{Institute of High Energy Physics, Chinese Academy of Sciences, Beijing}
\affiliation{Institute of High Energy Physics, Vienna}
\affiliation{Institute of High Energy Physics, Protvino}
\affiliation{Institute for Theoretical and Experimental Physics, Moscow}
\affiliation{J. Stefan Institute, Ljubljana}
\affiliation{Kanagawa University, Yokohama}
\affiliation{Korea University, Seoul}
\affiliation{Kyoto University, Kyoto}
\affiliation{Kyungpook National University, Taegu}
\affiliation{Ecole Polyt\'ecnique F\'ed\'erale Lausanne, EPFL, Lausanne}
\affiliation{University of Ljubljana, Ljubljana}
\affiliation{University of Maribor, Maribor}
\affiliation{University of Melbourne, School of Physics, Victoria 3010}
\affiliation{Nagoya University, Nagoya}
\affiliation{Nara Women's University, Nara}
\affiliation{National Central University, Chung-li}
\affiliation{National United University, Miao Li}
\affiliation{Department of Physics, National Taiwan University, Taipei}
\affiliation{H. Niewodniczanski Institute of Nuclear Physics, Krakow}
\affiliation{Nippon Dental University, Niigata}
\affiliation{Niigata University, Niigata}
\affiliation{University of Nova Gorica, Nova Gorica}
\affiliation{Osaka City University, Osaka}
\affiliation{Osaka University, Osaka}
\affiliation{Panjab University, Chandigarh}
\affiliation{Peking University, Beijing}
\affiliation{University of Pittsburgh, Pittsburgh, Pennsylvania 15260}
\affiliation{Princeton University, Princeton, New Jersey 08544}
\affiliation{RIKEN BNL Research Center, Upton, New York 11973}
\affiliation{Saga University, Saga}
\affiliation{University of Science and Technology of China, Hefei}
\affiliation{Seoul National University, Seoul}
\affiliation{Shinshu University, Nagano}
\affiliation{Sungkyunkwan University, Suwon}
\affiliation{University of Sydney, Sydney, New South Wales}
\affiliation{Tata Institute of Fundamental Research, Mumbai}
\affiliation{Toho University, Funabashi}
\affiliation{Tohoku Gakuin University, Tagajo}
\affiliation{Tohoku University, Sendai}
\affiliation{Department of Physics, University of Tokyo, Tokyo}
\affiliation{Tokyo Institute of Technology, Tokyo}
\affiliation{Tokyo Metropolitan University, Tokyo}
\affiliation{Tokyo University of Agriculture and Technology, Tokyo}
\affiliation{Toyama National College of Maritime Technology, Toyama}
\affiliation{Virginia Polytechnic Institute and State University, Blacksburg, Virginia 24061}
\affiliation{Yonsei University, Seoul}
  \author{K.~Abe}\affiliation{High Energy Accelerator Research Organization (KEK), Tsukuba} % KEK
  \author{I.~Adachi}\affiliation{High Energy Accelerator Research Organization (KEK), Tsukuba} % KEK
  \author{H.~Aihara}\affiliation{Department of Physics, University of Tokyo, Tokyo} % Tokyo
  \author{K.~Arinstein}\affiliation{Budker Institute of Nuclear Physics, Novosibirsk} % BINP
  \author{T.~Aso}\affiliation{Toyama National College of Maritime Technology, Toyama} % Toyama
  \author{V.~Aulchenko}\affiliation{Budker Institute of Nuclear Physics, Novosibirsk} % BINP
  \author{T.~Aushev}\affiliation{Ecole Polyt\'ecnique F\'ed\'erale Lausanne, EPFL, Lausanne}\affiliation{Institute for Theoretical and Experimental Physics, Moscow} % ITEP
  \author{T.~Aziz}\affiliation{Tata Institute of Fundamental Research, Mumbai} % Tata
  \author{S.~Bahinipati}\affiliation{University of Cincinnati, Cincinnati, Ohio 45221} % Cincinnati
  \author{A.~M.~Bakich}\affiliation{University of Sydney, Sydney, New South Wales} % Sydney
% ...
% Lausanne
\collaboration{The Belle Collaboration}
\noaffiliation

\begin{abstract}
We have measured branching fractions of hadronic $\tau$ decays involving 
an $\eta$ meson  using 485 fb$^{-1}$ of data collected with the Belle 
detector at the KEKB asymmetric-energy $e^+e^-$ collider. 
%A precise measurement of the following branching fractions has been
%performed: 
We obtain the following branching fractions:
${\cal B}(\tau^-\to K^- \eta \nu_{\tau})=(1.62\pm 0.05 \pm 0.09)\times 10^{-4}$, 
${\cal B}(\tau^-\to K^- \pi^0 \eta \nu_{\tau})  =(4.7\pm 1.1 \pm 0.4)\times 10^{-5}$, 
${\cal B}(\tau^-\to \pi^- \pi^0 \eta \nu_{\tau})=(1.39 \pm 0.03 \pm 0.07)
\times 10^{-3}$, and 
${\cal B}(\tau^-\to K^{*-} \eta \nu_{\tau})=(1.13\pm 0.19 \pm 0.07)\times 10^{-4}$ 
improving the accuracy compared to the best previous measurements by factors of six, eight, four and four, respectively.
\end{abstract}

%\pacs{13.65.+i, 13.25.Gv, 14.40.Gx}

\maketitle

%%%% >>>> keep the final version single-spaced
\tighten

{\renewcommand{\thefootnote}{\fnsymbol{footnote}}}
\setcounter{footnote}{0}

\section{INTRODUCTION}

 Hadronic decays of $\tau$ lepton provide a useful tool for 
studying QCD phenomena at low energy. Various decay modes 
including $\eta$ meson(s) are interesting for testing the
Wess-Zumino-Witten (WZW) anomaly~\cite{WZ,Witten}, chiral 
theory~\cite{Pich,Li}, and relations to $e^+e^-$ cross sections following
from the conservation of the vector current (CVC)~\cite{CVC}.

We measure the branching fractions of $\tau^- \to K^- \eta \nu_{\tau}$
(unless specified otherwise, charge conjugate decays are implied throughout the paper), $K^-
\pi^0 \eta \nu_{\tau}$, and $\pi^- \pi^0 \eta \nu_{\tau}$ decays, and that of
$\tau^-\to K^{^*-}(892) \eta \nu_{\tau}$; the latter is evaluated from the
corresponding $K^- \pi^0 \eta \nu_{\tau}$ measurement.
Studies of these modes have been reported 
by CLEO~\cite{CLEOpipi0etanu,CLEOketanu,CLEOkpi0etanu} and 
ALEPH~\cite{ALEPH}, however, most of the results are based on rather low 
statistics, which do not allow one to
discriminate between different theoretical predictions. 
We use a data sample with an integrated luminosity of 485 fb$^{-1}$, 
corresponding to production of 430 million $\tau$-pairs  
collected with the Belle detector at the KEKB asymmetric-energy
$e^+e^-$ collider~\cite{KEKB}.

The Belle detector is a large-solid-angle magnetic spectrometer that
consists of a silicon vertex detector (SVD),
a 50-layer central drift chamber (CDC), an array of
aerogel threshold Cherenkov counters (ACC),
a barrel-like arrangement of time-of-flight
scintillation counters (TOF), and an electromagnetic calorimeter
comprised of CsI(Tl) crystals (ECL) located inside
a superconducting solenoid coil that provides a 1.5~T
magnetic field.  An iron flux-return located outside 
the coil is instrumented to detect $K_L^0$ mesons and identify
muons (KLM).  The detector is described in detail elsewhere~\cite{BelleDet}.
Two inner detector configurations were used. A 2.0 cm radius beampipe
and a 3-layer silicon vertex detector were used for the first sample
of 144~fb$^{-1}$, while a 1.5 cm radius beampipe, a 4-layer
silicon detector and a small-cell inner drift chamber were used to record
the remaining 341~fb$^{-1}$\cite{svd2}.

In this analysis, we use a data sample 100 times larger than any of the
previous measurements. In addition, peaking backgrounds are estimated
precisely to decrease systematic uncertainties. 
%Thus, branching fractions for modes reported in this paper have the world's best accuracy. 

A study of resonance formation in the hadronic final states is in progress
and will be reported later.
  
\section{EVENT SELECTION}

 Candidate $e^+ e^- \to \tau^+\tau^-$ events are selected with the following common properties: 
\begin{center}
$e^+ e^- \to \tau_{\rm tag}^+ \tau_{\rm sig}^-$ \hspace*{9 cm} \\
\hspace*{1 cm} $\tau_{\rm sig}^- \to X\eta\nu_{\tau}$  ~~ and ~~ 
$\tau_{\rm tag}^+ \to (e/\mu)^+\nu_{l}\bar{\nu_{\tau}}$,
\end{center} 
where $X$ denotes $K^-$, $K^-\pi^0$, or $\pi^-\pi^0$. Candidate $\eta$ mesons 
are reconstructed through two decay modes: 
$\gamma\gamma$ with a  branching fraction of $39.39\pm 0.24\%$ or 
$\pi^+\pi^-\pi^0$ with a branching fraction of $22.68\pm 0.35\%$ 
with $\pi^0 \to \gamma\gamma$. 
In order to remove $q\overline{q}$ contamination, the tag-side $\tau$ 
is required to decay leptonically i.e.
$\tau^-\to\ell^-\nu_{\tau}\bar{\nu_{l}}$ $(\ell=e/\mu)$ (with a  branching 
fraction of $35.20 \pm 0.07\%$). All branching fractions are taken from
Ref.~\cite{PDG}. \\ 

\underline{$\tau^-\to K^-\eta\nu_{\tau}$ selection}

A candidate event is required to contain either two charged tracks 
with zero net charge and at least two $\gamma$'s in the 
$\eta\to\gamma\gamma$ case, or 
four charged tracks with zero net charge and at least two $\gamma$'s 
in the $\eta\to\pi^+\pi^-\pi^0$ $(\pi^0\to\gamma\gamma)$ case. 
A charged track should have transverse momentum, 
$p_t> 0.1$ GeV/c, and $-0.866 < \cos\theta < 0.956$ 
($\theta$ is the polar angle within the detector aperture) where $p_t$ and
$\theta$ are measured relative to the direction opposite to that of the incident
$e^+$ beam in the laboratory frame, and the
$\gamma$ should have energy $E_{\gamma} > 0.05$ GeV within the 
same polar angle fiducial region as above. 

To select a sample of $\tau$-pairs, the thrust vector and total energy 
in the center-of-mass system are required to satisfy $|V_{\rm thrust}| > 0.8$ 
and $3.0 < E^{\rm CM}_{\rm total} < 10.0$ GeV, respectively. 
Each event  is then subdivided perpendicularly to the thrust axis 
into two hemispheres, the signal and the tag. To reduce
$q\overline{q}$ contamination the effective mass of the particles on 
the tag side should satisfy the requirements, $M_{\rm tag} < m_{\tau}$ (1.78 GeV/c$^2$), 
while on the signal side the mass, $M_{\rm sig}$, should satisfy $0.70~ {\rm GeV/c}^2 < M_{\rm sig} < m_{\tau}$. 

To reconstruct an $\eta$ from $2\gamma$'s, two $\gamma$'s with 
$E_{\gamma} > 0.2$ GeV are required in the barrel region 
($-0.624 < \cos\theta < 0.833$) with at most one $\gamma$ 
with $0.05 < E_{\gamma} < 0.2$ GeV in the endcap region to allow for 
initial state radiation. 
In order to reduce incorrect combinations with a $\gamma$ from a
$\pi^0$ decay (denoted hereafter as $\gamma_{\pi^0}$), the $\eta$-candidate 
$\gamma$ ($\gamma_{\eta}$) 
should not form a $\pi^0$ mass with any other $\gamma$ ($\pi^0$ veto selection). 
In this analysis  the $\pi^0$ mass window is defined as 
0.105 GeV/$c^2~<~M_{\gamma\gamma}~<$ 0.165 GeV/$c^2$ 
(a $\pm 3 \sigma$ range in the detector resolution).
Particle identification (PID) uses a 
likelihood ratio, ${\cal P}_x$, for a charged particle of species $x$
($x=\mu,~e,~K,~\textrm{or}~\pi$).
${\cal P}_x$ is defined as ${\cal P}_x=L_x /\Sigma_y L_y$
(the sum runs over the relevant particle species), where $L_x$ is a likelihood based on 
the energy deposit and shower shape in the ECL, 
the momentum and $dE/dX$ measured by the CDC,
the particle range in the KLM, the light yield in the ACC, and particle's time-of-flight 
from the TOF counter~\cite{EMu}.
For the track on the tag side, ${\cal {P}}_{e} > 0.8$ 
for an $e$ candidate and ${\cal {P}}_{\mu} > 0.8$ for a $\mu$ candidate with $p > 0.7$ GeV/c. 
On the other hand, a kaon is identified as a track satisfying not only 
${\cal {P}}_{K} > 0.9$, but also ${\cal {P}}_{e} < 0.2$ in order to suppress 
two-photon events, with $p > 0.3$ GeV/c.

\begin{figure}[t]
 \begin{center}
  \includegraphics[keepaspectratio=true,height=50mm]{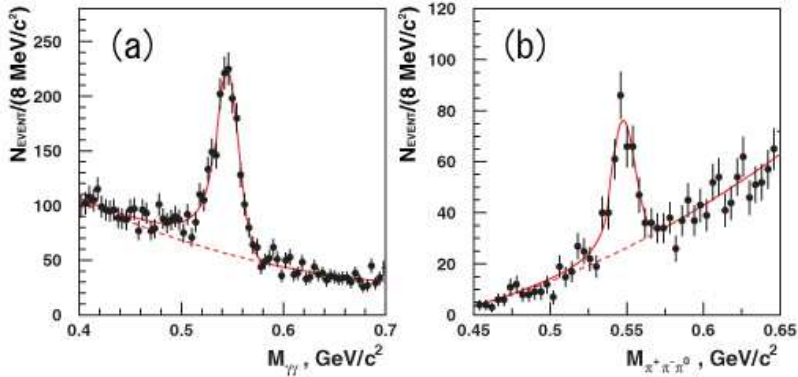}
 \end{center}
\caption{$M_{\gamma\gamma}$ distributions for $K^-\eta\nu_{\tau}$ selection in 
(a) $\eta\to\gamma\gamma$ and (b) $\eta\to\pi^-\pi^+\pi^0$ decays. 
Data are fit with a Crystal Ball function 
plus a second-order polynomial for the background (BG). 
The best fit result is indicated by the solid curve with the BG shown 
by the dashed curve.}
\label{fig:fitdata.eps}
\end{figure}

We also include correlations between tracks and $\gamma$'s.
The polar angle of the missing momentum that must be attributed to
neutrinos, should satisfy 
$-0.866 < \cos\theta(P_{\rm miss}) < 0.956$. 
The opening angle between the $K^-$ and $\eta$ satisfies the
requirement,
$\cos\theta(P^{\rm CM}_{K}; P^{\rm CM}_{\eta}) > 0.8$. The opening angle and 
energy of two $\gamma_{\eta}$'s should satisfy the following condition:
$0.5 < \cos\theta(P^{\rm CM}_{\gamma 1}; P^{\rm CM}_{\gamma 2}) < 0.96$ and  
$(E^{\rm CM}_{\gamma 1}-E^{\rm CM}_{\gamma 2})/(E^{\rm CM}_{\gamma 1}+
E^{\rm CM}_{\gamma 2}) < 0.8$, respectively.

%These selection criteria result in a detection efficiency of
%$\varepsilon = 0.94 \%$ including the branching fractions, 
%${\cal B}(\eta\to\gamma\gamma)$ and ${\cal B}(\tau\to l\nu_{\tau}\bar{\nu})$.  
The $\gamma\gamma$ mass distribution obtained after these requirements is shown in Fig.~1 (a). 
 
In the case of $\eta\to\pi^+\pi^-\pi^0$ reconstruction, 
candidate events should have two additional charged tracks compared
to the $\eta\to\gamma\gamma$ sample, 
but two of the $\gamma$'s have to form a $\pi^0$ instead of an $\eta$. 
While this mode provides one more constraint compared to those in 
the previous case improving the background rejection, the
higher multiplicity reduces the detection efficiency. 

The selection criteria different from those in the previous case 
are indicated below. 
%The thrust cut is changed a bit to optimize a signal to background ratio, 
%$\mid V_{thrust}\mid > 0.9$ from 0.8.
$\pi^+$ and $\pi^-$ candidates are required to have 
$0.3~{\rm GeV}/c~<~p~<~2.0~{\rm GeV}/c$ 
and to be inconsistent with the $e$ hypothesis (${\cal P}_{e} < 0.2$) to reject 
the two-photon background. 
The photons used to form $\pi^0$-candidates are required to have $E_{\gamma} > 0.1$ GeV. 
In order to remove the contributions from higher mass states such as
decays of an $\omega$ meson, the condition
$M_{\pi^+\pi^-\pi^0} < 0.7$ GeV/$c^2$ is imposed. 
The condition on the polar angle of the missing momentum is the same 
as in the previous case while the $\pi^0$ momentum should be 
$P^{\rm CM}_{\pi^0} > 0.5$ GeV/$c$. 

The resulting $\pi^+\pi^-\pi^0$ mass distribution is shown in Fig.1~(b).\\
% and the detection efficiency $\varepsilon = 0.16 \%$ is obtained from Monte 
%Carlo simulation. \\

\underline{Selection of $\tau^-\to K^- \pi^0 \eta \nu_{\tau}$ and 
$\pi^-\pi^0\eta\nu_{\tau}$} 

\begin{figure}[t]
 \begin{center}
  \includegraphics[keepaspectratio=true,height=50mm]{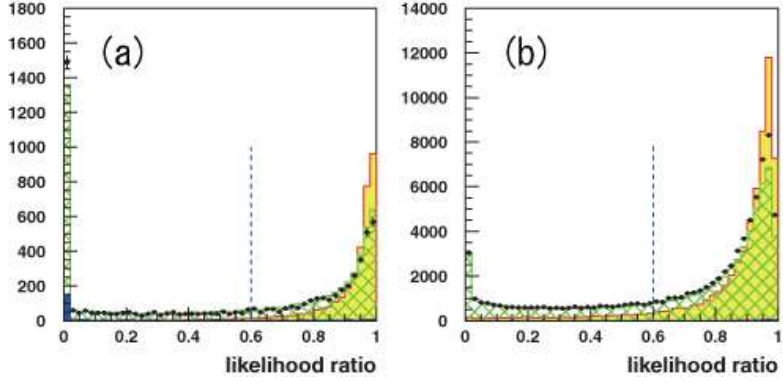}
 \end{center}
\caption{Distribution of likelihood ratios for (a) $K^-\pi^0\eta\nu_{\tau}$ and 
(b) $\pi^-\pi^0\eta\nu_{\tau}$ candidates. The points with error bars are the data. The yellow, green
 hatched, and blue histograms indicate the signal, $\tau\tau$ background, and
 $q\overline{q}$ background MC distributions,
 respectively. The MC histograms are normalized to the
 integrated luminosity of the data. For demonstration purposes, the
 branching fractions of signal MC decays are set to $10^{-3}$ and
 $10^{-2}$ for (a) and (b), respectively. The dashed vertical line show
 the likelihood ratio requirement.}
\label{fig:Likelihood}
\end{figure}

For these $\tau$ decay modes, we use only $\eta\to\gamma\gamma$ to reconstruct the $\eta$, because of 
the small detection efficiency for $\eta\to \pi^+\pi^-\pi^0$. 
Correspondingly, a candidate event should contain two charged tracks
and at least four $\gamma$'s. 

The selection criteria different from those for $K^-\eta\nu_{\tau}$ are listed
below. The total momentum on the signal side is required to satisfy 
$\sum P^{\rm CM}_{\rm sig} > 2.5$ GeV/$c$;
%for an additional $\pi^0$ we require 
two additional $\gamma$'s are required to lie in the barrel
region on the signal side that form the $\pi^0$ mass; 
a condition on cosine of the opening angle between the 
missing momentum and the direction of the thrust axis
pointing to the signal side is imposed: 
$\cos\theta(P^{\rm CM}_{\rm miss}; P^{\rm CM}_{\rm thrust}) < -0.6$. 
For the $\pi^-\pi^0\eta\nu_{\tau}$ mode, $\pi^-$ candidates should have
${\cal P}_{K} < 0.2$ and ${\cal P}_e < 0.2$. 

To further suppress backgrounds, we apply a likelihood selection using 
seven variables, such as 
$|V_{\rm thrust}|$, $P^{\rm CM}_{\eta}$, 
$M^2_{\rm miss}$ (missing mass squared), 
$P^{\rm CM}_{\pi^0}$, $E^{\rm CM}_{\gamma_{\eta}}$, 
$\sum P^{\rm CM}_{\rm sig}$, and 
$\cos\theta(P^{\rm CM}_{K/\pi}; P^{\rm CM}_{\eta})$. The
$\tau\tau$ background MC is used for the background likelihood while the
distributions from $K^-\pi^0\eta\nu$ and $\pi^-\pi^0\eta\nu$ MC are used
for the signal likelihood in each case.
The resulting likelihood ratios defined for $K^-\pi^0\eta\nu_{\tau}$ and 
$\pi^-\pi^0\eta\nu_{\tau}$ are shown in Figs.2~(a) and (b), respectively.
About half of the background is removed while 
93\% and 90\% of the signal is retained for each of the respective modes. 

The obtained $M_{\gamma\gamma}$ distributions are shown in Figs.~3 (a) 
and (b) for the $K^-\pi^0\eta\nu_{\tau}$ and $\pi^-\pi^0\eta\nu_{\tau}$ modes,
respectively.
% and the corresponding detection efficiencies are $\varepsilon = 0.35 \%$ and $0.47 \%$. 

\begin{figure}[t]
 \begin{center}
  \includegraphics[keepaspectratio=true,height=50mm]{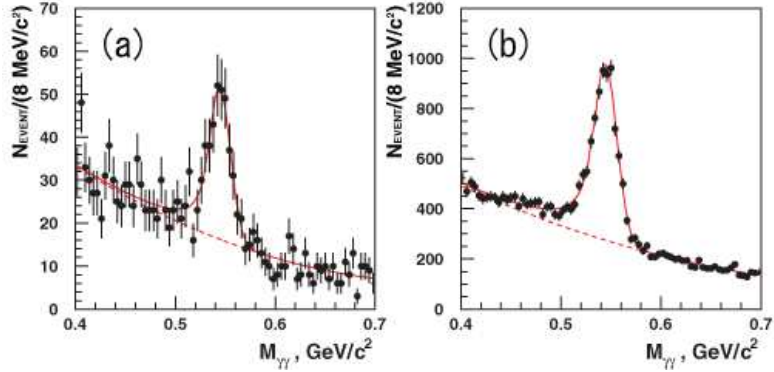}
 \end{center}
\caption{$M_{\gamma\gamma}$ distributions for (a) $K^-\pi^0\eta\nu_{\tau}$ and 
(b) $\pi^-\pi^0\eta\nu_{\tau}$ candidates. 
Data are fit with a Crystal Ball function plus a second-order polynomial 
for the background (BG). 
The result of the best fit is indicated by the solid curve with the BG 
shown by the dashed curve.}
\label{fig:fig2.eps}
\end{figure}

\section{BRANCHING FRACTIONS FOR $K^-\eta\nu_{\tau}$, $K^-\pi^0\eta\nu_{\tau}$ and
 $\pi^-\pi^0\eta\nu_{\tau}$ DECAYS} 

To determine the signal yields we determine the number of $\eta$'s and
then subtract cross-feeds. In order to extract the number of $\eta$'s from the resulting 
$M_{\gamma\gamma}$ distributions, 
Figs.~1 and 3, we use the Crystal Ball function~\cite{CB} to represent 
the $\eta$ signal and a second-order polynomial for the 
background distribution. 
The result of the fits is indicated by the solid curve in the corresponding 
figures. The best fits give an $\eta$ mass, 
$m_{\eta}$, of $0.545\pm 0.001$ GeV/$c^2$, in good agreement with 
the PDG value of $0.54751\pm 0.00018$ GeV/$c^2$, and 
a resolution $\sigma_{m_{\eta}} = 0.012\pm 0.002$ GeV/$c^2$ for the
three different decay modes in the $\eta\to\gamma\gamma$ case and 
$m_{\eta} = 0.5474\pm 0.0007$ GeV/$c^2$ with 
$\sigma_{m_{\eta}} = 0.0075\pm 0.0004$ GeV/c$^2$ 
for the $K^-\eta\nu_{\tau}$ mode in the $\eta\to\pi^+\pi^-\pi^0$ case.  

The $\eta$ yields obtained from the fits are 
$N_{K^-\eta\nu_{\tau}}=1387\pm 43$, $N_{K^-\pi^0\eta\nu_{\tau}}=270\pm 33$, and 
$N_{\pi^-\pi^0\eta\nu_{\tau}}=5959\pm 105$ events for the 
$K^-\eta\nu_{\tau}$, $K^-\pi^0\eta\nu_{\tau}$, and 
$\pi^-\pi^0\eta\nu_{\tau}$ modes, respectively, in 
the $\eta\to\gamma\gamma$ case. The yield for the $\eta\to\pi^+\pi^-\pi^0$ 
case is $N_{K^-\eta\nu_{\tau}; \eta\to 3\pi}=241\pm 21$ events. 
The $\eta$ signal region is defined as 
$0.48~{\rm GeV}/c^2~<~M_{\gamma\gamma}~<~0.58~{\rm GeV}/c^2$ in 
the $\eta\to\gamma\gamma$ case while it is 
$0.52~{\rm GeV}/c^2~<~M_{\pi^+\pi^-\pi^0}~<~0.58~{\rm GeV}/c^2$
in the $\eta\to\pi^+\pi^-\pi^0$ case. The detection efficiencies are
estimated with MC simulation in the same
manner as the detection of $\eta$ yields. The corresponding efficiencies
are $\varepsilon = 0.94~\%$, $0.35~\%$, $0.47~\%$, and $0.16~\%$,
respectively, and include the intermediate branching fractions
such as ${\cal B}(\eta\to\gamma\gamma)$, 
${\cal B}(\eta\to\pi^+\pi^-\pi^0)$, and
${\cal B}(\tau^-\to l^-\nu_{\tau}\bar{\nu_{l}})$.

These event yields include backgrounds classified into three categories.
One is due to cross-feed effects between the three signal modes, which is 
taken into account by solving the following system of equations:
\begin{equation}\label{eq1}
 N_{K^- \eta \nu_{\tau}}=2N_{\tau\tau}({\cal B}(K^-\eta\nu_{\tau}) \cdot \epsilon_1^1 + 
{\cal B}(K^-\pi^0\eta\nu_{\tau}) \cdot
 \epsilon_2^1 + {\cal B}(\pi^-\pi^0\eta\nu_{\tau}) \cdot \epsilon_3^1),
\end{equation}
\begin{equation}
N_{K^- \pi^0 \eta \nu_{\tau}}=2N_{\tau\tau}({\cal B}(K^-\eta\nu_{\tau}) \cdot \epsilon_1^2 + 
{\cal B}(K^-\pi^0\eta\nu_{\tau}) \cdot
 \epsilon_2^2 + {\cal B}(\pi^-\pi^0\eta\nu_{\tau}) \cdot \epsilon_3^2),
\end{equation}
\begin{equation}\label{eq3}
N_{\pi^- \pi^0 \eta \nu_{\tau}}=2N_{\tau\tau}({\cal B}(K^-\eta\nu_{\tau}) \cdot \epsilon_1^3 
+ {\cal B}(K^-\pi^0\eta\nu_{\tau}) \cdot
 \epsilon_2^3 + {\cal B}(\pi^-\pi^0\eta\nu_{\tau}) \cdot \epsilon_3^3),
\end{equation}
where ${\cal B}(K^-\eta\nu_{\tau})$, ${\cal B}(K^-\pi^0\eta\nu_{\tau})$, and 
${\cal B}(\pi^-\pi^0\eta\nu_{\tau})$ are the branching fractions of the 
respective modes and $N_{\tau\tau}$ is the total number of
$\tau$ pairs produced. 
Here $\varepsilon_j^i$ is the detection efficiency in each
case and $j$ ($i$) indicates the decay sample (selection criteria).

The second type of background is from decay modes of 
the $\tau$-lepton itself, 
such as $\pi^-\pi^0\pi^0\eta\nu_{\tau}$ and $\pi^-\pi^+\pi^-\eta\nu_{\tau}$. 
These backgrounds are estimated by using a $\tau\tau$ MC simulation with 
branching fractions taken from \cite{CLEO6pion}. 
The estimated background is included in the `other' category in Table~\ref{Table:event}.
The contribution of this type of background is negligible for 
the $K^-\eta\nu_{\tau}$ and $K^-\pi^0\eta\nu_{\tau}$
modes, and is smaller than the statistical uncertainty in the  
$\eta$ yield obtained above for the $\pi^-\pi^0\eta\nu_{\tau}$ mode.
The contamination from $\pi^-\eta\nu_{\tau}$ decay should also be
considered. This decay is strongly suppressed since it violates
$G$-parity and proceeds via a second-class current. Its branching fraction is predicted to be $10^{-5}$. Therefore, its
contribution to each mode is negligible.

The last background category is $e^+e^- \to q\bar{q}$, which is
estimated from MC simulation. The MC was tuned
beforehand and validated with a $q\bar{q}$ enriched sample,
which was produced with some variations of
the event selection criteria. The $M_{\rm sig}$ cut is removed and 
the condition
$M_{\rm tag}>m_{\tau}$ is implemented on the tag side. In addition, the
PID requirement for the charged track on the tag side was reversed
(i.e. ${\cal P}_e<0.8$ and ${\cal P}_\mu<0.8$). The $\eta$ yield of 
MC was then tuned to be consistent with that of data.
The $q\bar{q}$ contributions are 2-4 \% of the $\eta$ yields for the 
$K^-\eta\nu_{\tau}$ and $\pi^-\pi^0\eta\nu_{\tau}$
modes and 10\% for the $K^-\pi^0\eta\nu_{\tau}$. They also are summarized in 
Table~\ref{Table:event}.

\begin{table}
\caption {Raw $\eta$ yields for the four selected $\tau$ decay modes.
$N_{\eta}$ is the total number of $\eta$ events detected, which include
 cross-feed from two other modes.
The contributions of $q\overline{q}$ and `other', mostly 
$\pi^-\pi^0\pi^0\eta\nu_{\tau}$ and $\pi^-\pi^+\pi^-\eta\nu_{\tau}$, are also included.}
 \label{Table:event}
 \begin{center}
  \begin{tabular}{|c|c|c|c|c|c|c|c|}
   \hline
 Mode                        & $N_{\eta}$ && $K\eta\nu_{\tau}$ &
   $K\pi^0\eta\nu_{\tau}$ & $\pi\pi^0\eta\nu_{\tau}$ & Other & $q\bar{q}$ \\
   \hline
$K^{-}\eta\nu_{\tau}$ ($\eta \to \gamma\gamma$)           &$1387\pm43$
   &&$-$ &$15.1 \pm 3.8$&$18.0\pm 1.0$& $1.1\pm0.2$ & $30.6\pm15.6$\\
$K^{-}\pi^0 \eta \nu_{\tau}$     &$270\pm33$&&$16.0\pm 0.9$&
   $-$&$85.3\pm 4.6$&$1.2\pm0.4$ & $27.0\pm8.5$\\
$\pi^{-}\pi^0 \eta \nu_{\tau}$    & $5959\pm105$&&$2.4\pm
   0.1$&$9.4\pm 2.4$&$-$ &$71.6\pm20$ & $212\pm29$\\
$K^{-}\eta\nu_{\tau}$ ($\eta \to \pi^+\pi^-\pi^0$) &$241\pm21$ &&$-$ &$3.3\pm0.8$ &$5.8\pm1.3$ & $<1.18$&$9.1\pm2.2$ \\
%\textbf{$K^{*-} \to K^-\pi^0$ detection} & & & & & & \\
%$K^{*-}\eta\nu_{\tau}$ &$119\pm17$ &$6.5\pm2.3$ & & & & $2.3\pm0.9$ \\
   \hline
  \end{tabular}
 \end{center}
\end{table}

After subtracting the `other' and $q\overline{q}$ contributions from 
the individual $\eta$ yields, we solve the system of 
equations to obtain the branching fractions for three decay modes. 
They are $(1.62\pm 0.05)\times 10^{-4}$, $(4.7\pm 1.1)\times 10^{-5}$, and 
$(1.39\pm 0.03)\times 10^{-3}$ for $K^-\eta\nu_{\tau}$, $K^-\pi^0\eta\nu_{\tau}$, and 
$\pi^-\pi^0\eta\nu_{\tau}$, respectively. The cross-feed yields obtained
are also listed in Table \ref{Table:event}. 
The number of cross-feed events for $K^-\eta\nu_{\tau}$ in the 
$\eta\to\pi^+\pi^-\pi^0$ case 
is evaluated using the above branching fractions, obtained in 
the $\eta\to\gamma\gamma$ case. 
The branching fraction for $K^-\eta\nu_{\tau}$ in this case is 
$(1.65\pm 0.16)\times 10^{-4}$.\\
 
The systematic uncertainties are estimated as follows: the estimation of
peaking backgrounds provides sizable uncertainties only in case of the
$\pi^-\pi^0\eta\nu_{\tau}$ (3.3 $\%$) and $q\overline{q}$ (6.0 $\%$) contributions to the $K^-\pi^0\eta\nu_{\tau}$
mode. As for the $q\bar{q}$ estimation, due to the finite statistics of
$q\bar{q}$ enriched sample, the uncertainties of 26 $\%$, 19 $\%$, and
9.6 $\%$ for $K^-\eta\nu_{\tau}$, $K^-\pi^0\eta\nu_{\tau}$, and
$\pi^-\pi^0\eta\nu_{\tau}$ decays arise from tuning,
respectively. The errors in the $q\bar{q}$ background estimations in Table
\ref{Table:event} come from this uncertainty and the statistical
uncertainty in the $q\bar{q}$ MC, and are treated as systematic
uncertainties. The uncertainties in the peaking backgrounds in all other
cases are rather small. Uncertainties in the
PID efficiency and fake rate are evaluated to be 2-3 $\%$ for kaon ID, 1
$\%$ for $\pi$ ID and
around 2.5 $\%$ for the lepton ID; these values are obtained by
averaging the estimated
uncertainties depending on momentum and polar angle of each charged
track. For the $\pi^0$ veto selection, the efficiency was compared
between data and MC with a sample in which the PID of
the charged track on the tag side was reversed from the usual
$\pi^-\pi^0\eta\nu_{\tau}$ selection. The efficiencies were consistent.
Therefore, 2.8 $\%$ of the statistical uncertainty from this comparison was
counted as systematic uncertainty.
Other systematic uncertainties are summarized in Table
\ref{Table:sysKetanu}. The total systematic uncertainties are 5.6 $\%$, 8.9
$\%$, 5.0 $\%$, and 6.3 $\%$ for the $K^-\eta\nu_{\tau} \
(\eta\to\gamma\gamma)$, $K^-\pi^0\eta\nu_{\tau}$,
$\pi^-\pi^0\eta\nu_{\tau}$, and $K^-\eta\nu_{\tau} \
(\eta\to\pi^+\pi^-\pi^0)$ modes, respectively. 

\begin{table}
 \caption {Summary of systematic uncertainties in each mode ($\%$)}
 \label{Table:sysKetanu}
 \begin{center}
  \begin{tabular}{|c|c|c|c|c|c|}
   \hline
    Mode                   & $K^-\eta\nu_{\tau}$ & $K^-\pi^0\eta\nu_{\tau}$  & $\pi^-\pi^0\eta\nu_{\tau}$ & $K^-\eta\nu_{\tau}$ & $K^{*-}\eta\nu_{\tau}$ \\
    $\eta$ detection       & $\eta \to \gamma\gamma$ & $\eta \to \gamma\gamma$  & $\eta \to \gamma\gamma$ & $\eta \to \pi^+\pi^-\pi^0$ & $\eta \to \gamma\gamma$ \\
   \hline
Estimation of $K^-\eta\nu_{\tau}$          &   $-$   & 0.6 & $1.8\times 10^{-3}$ & $-$ & $-$\\
Estimation of $K^-\pi^0\eta\nu_{\tau}$     &  0.3 & $-$ & $4.2\times 10^{-2}$ & 0.4 & $-$ \\
Estimation of $\pi^-\pi^0\eta\nu_{\tau}$   &  $7.5 \times 10^{-2}$ & 3.3 & $-$ & 0.1 & $-$ \\
Estimation of $\pi^-\pi^0\pi^0\eta\nu_{\tau}$   &  $-$ & $-$ & 0.4 & $-$ & $-$\\
%Estimation of $K^{*-}\nu_{\tau}$         &  $-$ & $-$ & $-$ & $-$ & 0.94 \\
Estimation of $q\bar{q}$          &  1.5 & 6.0 & 0.5 & 1.5 & 2.4  \\
Particle ID ($K/\pi$)              &  3.3 & 2.2 & 1.0 & 2.8 & 2.2 \\
Particle ID (Lepton)              &  2.3 & 2.8 & 2.6 & 2.6 & 2.6 \\
Track finding                          &  1.3 & 1.3 & 1.3 & 3.3 & 1.3 \\
Luminosity measurement            &  1.6 & 1.6 & 1.6 & 1.6 & 1.6 \\
$\pi^0$ detection                 &  $-$ & 2.0 & 2.0 & 2.0 & 2.0 \\
$\pi^0$ veto                      &  2.8 & 2.8 & 2.8 & $-$ & 2.8 \\
Signal MC                         &  0.5 & 1.7 & 0.7 & 1.3 & 1.7 \\
${\cal B}(\eta \to \pi^+\pi^-\pi^0)$ & $-$ & $-$ & $-$ & 1.6 &$-$ \\
   \hline
Total                             &  5.6 & 8.9 & 5.0 & 6.3 & 6.0 \\
   \hline
  \end{tabular}
 \end{center}
\end{table}

Taking the systematic uncertainties into account we obtain the following branching fractions:
\begin{eqnarray}
{\cal B}(\tau^-\to K^-\eta\nu_{\tau}) &=& (1.62\pm 0.05\pm 0.09)\times 10^{-4} ~~~~~~ 
 \textrm{for} \ \eta\to\gamma\gamma, \nonumber\\
&=& (1.65\pm 0.16\pm 0.10)\times 10^{-4} ~~~~~~ \textrm{for} \ \eta\to\pi^+\pi^-\pi^0,
\nonumber\\
{\cal B}(\tau^-\to K^-\pi^0\eta\nu_{\tau}) &=& (4.7~\pm 1.1\pm 0.4)\times 10^{-5},
\nonumber\\
{\cal B}(\tau^-\to \pi^-\pi^0\eta\nu_{\tau}) &=& (1.39\pm 0.03\pm 0.07)\times 10^{-3}.
\nonumber
\end{eqnarray}
By combining two measurements for $\tau^-\to K^-\eta\nu_{\tau}$ decay, we obtain: 
\begin{eqnarray}
{\cal B}(\tau^-\to K^-\eta\nu_{\tau}) &=& (1.62\pm 0.05 \pm 0.09)\times 10^{-4}. \nonumber
\end{eqnarray}

\section{BRANCHING FRACTION FOR $K^{*-}(892)\eta\nu_{\tau}$}
\begin{figure}[t]
 \begin{center}
  \includegraphics[keepaspectratio=true,height=100mm]{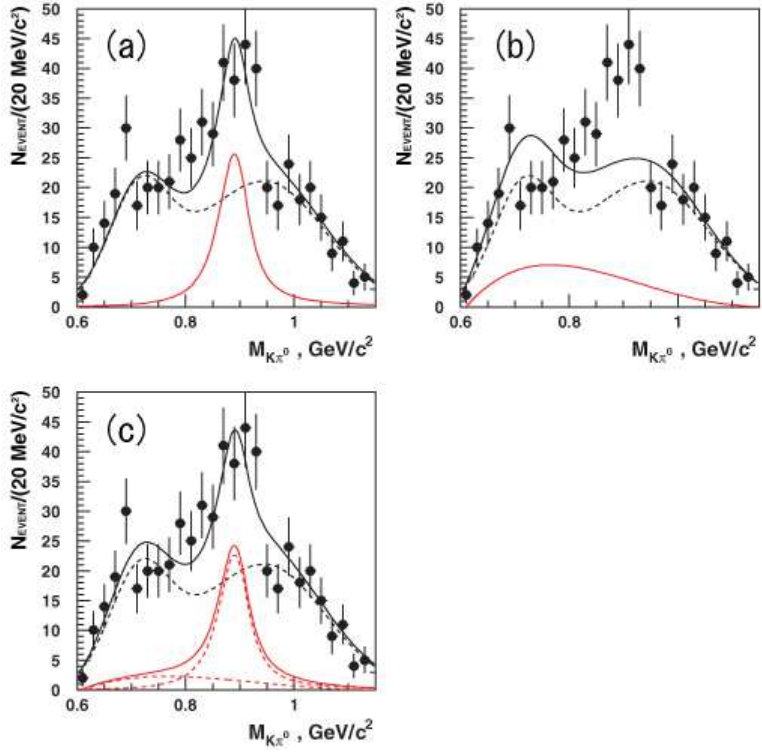}
 \end{center}
\caption{The $K^-\pi^0$ invariant mass distribution for 
the $K^-\pi^0\eta\nu_{\tau}$ events. 
Data are fitted with a Breit-Wigner (BW) function in (a), and the best 
fit is indicated by the solid 
curve while the continuum component evaluated from sideband regions is 
shown by the dashed curve. 
The fit gives $N_{K^{*-}\eta\nu_{\tau}}=119\pm 19$ events with 
$\chi^2/(\textrm{d.o.f.}=27) = 1.15$ 
(the probability to obtain this result is 0.265). 
(b) and (c) show the results of similar fit with a phase space distribution
 only and a BW plus a phase space distribution. 
In these cases the results are $N_{K^{*-}\eta\nu_{\tau}}=102\pm 21$ events 
with $\chi^2/(\textrm{d.o.f.}=27) = 2.09$ 
(the probability is 0.0008) for the phase-space distribution, or 
$N_{K^{*-}\eta\nu_{\tau}}=105\pm 20$ and $N_{K^-\pi^0\eta\nu_{\tau}}=33\pm 30$ events with 
$\chi^2/(\textrm{d.o.f.}=26) = 1.13$ (the probability is 0.294).}
\label{fig:fit_final.eps}
\end{figure}

From the $K^-\pi^0\eta\nu_{\tau}$ samples within the $\eta$ mass 
region, $0.50~{\rm GeV}/c^2~<~M_{\gamma\gamma}~<~0.58~{\rm GeV}/c^2$, 
we extract a branching fraction for $\tau^-\to K^{*-}(892)\eta\nu_{\tau}$
decay through the $K^{*-}(892)\to K^-\pi^0$ mode 
$({\cal B}(K^{*-}(892)\to K^-\pi^0)=1/3)$.
 
The distribution of $K^-\pi^0$ invariant mass, $M_{K^-\pi^0}$,  
for the selected samples is shown in Fig.~\ref{fig:fit_final.eps}.
The $\tau^- \to \pi^-\pi^0\eta\nu_{\tau}$ background, which cannot be
neglected as shown in Table \ref{Table:event}, is estimated by using MC
simulation with the branching fraction measured in this paper. The other
backgrounds are estimated from two sidebands of the 
$M_{\gamma\gamma}$ distribution: 
$0.43~{\rm GeV}/c^2~<~M_{\gamma\gamma}~<~0.48~{\rm GeV}/c^2$ 
and $0.60~{\rm GeV}/c^2~<~M_{\gamma\gamma}~<~0.65~{\rm GeV}/c^2$. 
The background distribution is indicated by the dashed curve.
The peculiar shape of the expected background is due to two components
contributing to it: that of $\pi^-\pi^0\eta\nu_{\tau}$ at high mass
and the one from all other $\tau$ decays at low mass.
A clear excess above the background is seen around 0.9 GeV/c$^2$, 
suggesting $K^{*-}(892)$ resonance formation. 
Three types of signal functions have been tested. A Breit-Wigner (BW) 
function whose mass and width are set to those of the $K^{*-}(892)$ 
is fitted to data in Fig.~\ref{fig:fit_final.eps}~(a).
A fit with the hypothesis that the excess signal events are due to 
a non-resonant, but $V-A$ pure phase space process is shown in
Fig.~\ref{fig:fit_final.eps}~(b). Figure~\ref{fig:fit_final.eps}~(c) shows
the result of a fit with a BW plus a phase space function as a signal. 
In each case, the signal function is smeared to take into account the 
detector resolution.
The BW function describes the data well in
Fig.~\ref{fig:fit_final.eps}~(a) and the resulting number of 
$K^{*-}(892)$ events is
$N_{K^{*-}\eta\nu_{\tau}} = 119\pm 19$. 
In contrast, the fit shown in Fig.~\ref{fig:fit_final.eps}~(b) is in complete 
disagreement with the data. 
Although inclusion of a small phase space component in addition to the 
dominant $K^{*-}$ resonance component also represents data 
well as shown in Fig.~\ref{fig:fit_final.eps}~(c), 
the resulting magnitude of the phase space component is consistent with 
zero within errors. 
Therefore, we conclude that all excess events are produced via  
the $K^{*-}(892)$ resonance. 

The detection efficiency is evaluated by MC simulation to be 
$\varepsilon = 0.12\%$, including the branching fractions of 
${\cal B}(K^{*-}(892)\to K^-\pi^0)$, ${\cal B}(\eta\to\gamma\gamma)$, and 
${\cal B}(\tau \to \ell\nu_{\tau}\overline{\nu})$. 

Since the requirement of $K^{*-}(892)$ formation in this case is a rather
strong constraint, no significant peaking background 
contribution from other $\tau$ decays is found. 
Therefore, the systematic uncertainty in the $K^{*-}(892)\eta\nu_{\tau}$
mode is small as compared with the
$K^-\pi^0\eta\nu_{\tau}$ analysis. Other 
sources of systematic uncertainties are the same as those of 
$K^-\pi^0\eta\nu_{\tau}$, except for the background contamination; a total
systematic uncertainty of 6.0 $\%$ is obtained with details
summarized in Table~\ref{Table:sysKetanu}. 
Finally, we obtain the following branching fraction: 
\begin{eqnarray}
{\cal B}(\tau^-\to K^{*-}\eta\nu_{\tau}) = (1.13\pm 0.19\pm 0.07) \times 10^{-4}.  
\end{eqnarray}

\section{RESULTS and DISCUSSION}

\begin{table}[t]
 \caption {Comparison of our measurement with previous results}
 \label{Table:Comparison}
 \begin{center}
  \begin{tabular}{|c|c|c|c|c|c|}
     \hline
  Mode & ${\cal B}$ in this analysis & Previous ${\cal B}$ & Reference \\
     \hline
$\tau^- \to K^-\eta\nu_{\tau}$  & $(1.62\pm0.05\pm0.09)\times 10^{-4}$ &
   $(2.6\pm0.5\pm0.5)\times 10^{-4}$ &CLEO~\cite{CLEOketanu}\\
 & & $(2.9\pm1.3\pm 0.7)\times 10^{-4}$ &ALEPH~\cite{ALEPH}\\
     \hline
$\tau^- \to K^-\pi^0\eta\nu_{\tau}$  & $(4.7\pm1.1\pm0.4)\times 10^{-5}$ & $(17.7\pm5.6\pm7.1)\times 10^{-5}$&CLEO~\cite{CLEOkpi0etanu}\\
     \hline
$\tau^- \to \pi^-\pi^0\eta\nu_{\tau}$  & $(1.39\pm 0.03\pm 0.07)\times 10^{-3}$
   & $(1.7\pm0.2\pm 0.2)\times 10^{-3}$&CLEO~\cite{CLEOpipi0etanu} \\
                        & & $(1.8\pm0.4\pm 0.2)\times 10^{-3}$&ALEPH~\cite{ALEPH} \\
     \hline
$\tau^- \to K^{*-}\eta\nu_{\tau}$  & $(1.13\pm 0.19\pm 0.07)\times 10^{-4}$ &$(2.90\pm 0.80\pm 0.42)\times 10^{-4}$&CLEO~\cite{CLEOkpi0etanu} \\
     \hline
  \end{tabular}
 \end{center}
\end{table}

We have obtained branching fractions for four different decay modes 
based on a high-statistics data sample of 430 million $\tau$-pairs collected 
with the  Belle detector:
\begin{eqnarray}
{\cal B}(\tau^-\to K^-\eta\nu_{\tau}) &=& (1.62\pm 0.05\pm0.09)\times 10^{-4}, \nonumber\\
{\cal B}(\tau^-\to K^-\pi^0\eta\nu_{\tau}) &=& (4.7\pm 1.1\pm 0.4)\times 10^{-5},\nonumber\\
{\cal B}(\tau^-\to \pi^-\pi^0\eta\nu_{\tau}) &=& (1.39\pm 0.03\pm 0.07)\times 10^{-3}, \nonumber\\
{\cal B}(\tau^-\to K^{*-}\eta\nu_{\tau}) &=& (1.13\pm 0.19\pm 0.07) \times 10^{-4}, \nonumber  
\end{eqnarray}
where the first error is  statistical and the second is systematic.
% except for the $K^-\eta\nu_{\tau}$. 
%The $B(\tau^-\to K^-\eta\nu_{\tau})$ is obtained by combining two results in 
%$\eta\to\gamma\gamma$ and $\eta\to\pi^-\pi^+\pi^0$ cases, where the common systematic 
%uncertainties are treated separately from the uncorrelated errors in a process of 
%adding the statistical and systematic uncertainties, in quadrature. 

Compared to previous experiments, we have improved not only 
the statistical uncertainties, but also evaluated 
reliably the background contamination. 
In Table \ref{Table:Comparison} our results are compared to those previously
obtained by the
CLEO~\cite{CLEOpipi0etanu,CLEOketanu,CLEOkpi0etanu} and
ALEPH~\cite{ALEPH} collaborations. 
Our measurement improves the uncertainties in the 
branching fractions by a factor of six ($K^-\eta\nu_{\tau}$), eight 
($K^-\pi^0\eta\nu_{\tau}$), four ($\pi^-\pi^0\eta\nu_{\tau}$), and four ($K^{*-}\eta\nu_{\tau}$)
compared to the most precise determinations from CLEO. In addition, the high 
statistics of our experiment allows much more reliable estimation
of various backgrounds including the peaking one.
The relatively poor statistics of previous measurements imposed some
limitations on BG estimations \footnote{For instance, for the
$K^-\pi^0\eta\nu_{\tau}$ measurement 
the cross-feed from $\pi^-\pi^0\eta\nu_{\tau}$ 
decay was not considered, although it gives the largest contribution 
as seen in Table \ref{Table:event}.}. It is also noteworthy that in all cases 
the central value of our measurement is lower than that
of the other measurements. This may be due to the underestimation of 
backgrounds in the previous experiments. The improved accuracy in the
branching fractions of the decay modes reported here is important for future
searches for the second-class current $\tau^- \to\pi^-\eta\nu_{\tau}$ decay.
%We can thus infer why the previous experiments result 
%in different values from ours, 
%even though their uncertainties are large. 

%Though TAUOLA MC program needs more tunes for reproducing observations, 
%specifically, for the $\tau$'s decay-modes involving Kaon(s), it does quite well 
%in reproducing decay-modes with multi-pion final states:
\begin{figure}[t]
 \begin{center}
  \includegraphics[keepaspectratio=true,height=100mm]{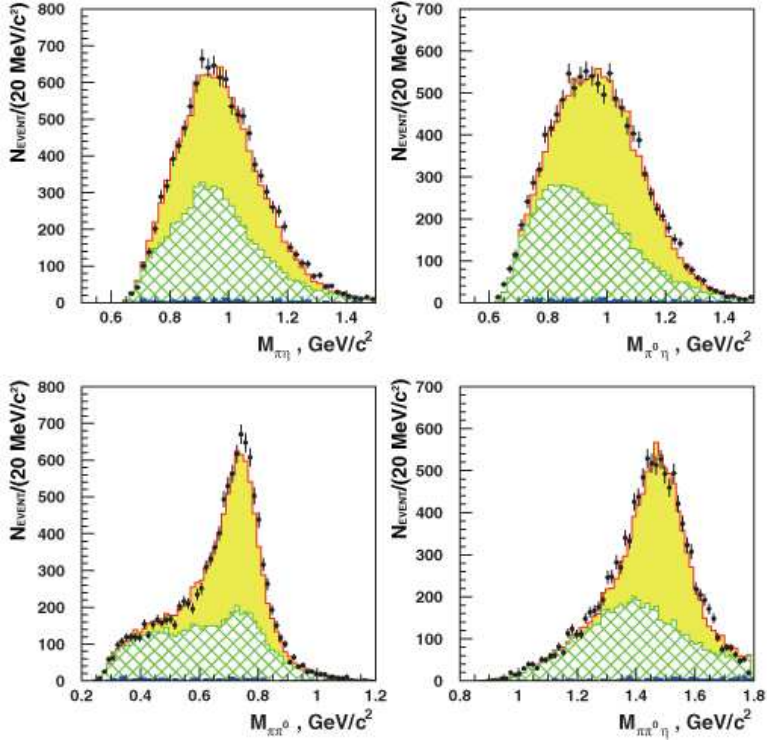}
 \end{center}
\caption{The invariant mass distributions of various combinations of
final state particles for
 $\tau^-\to\pi^-\pi^0\eta\nu_{\tau}$ decay. The points with error bars
 are the data. The yellow, green
 hatched, and blue histograms indicate the signal, $\tau\tau$
 background, and $q\overline{q}$ background MC distributions,
 respectively. The dominant backgrounds come from $\tau\tau$ events. The
 $q\overline{q}$ background is strongly suppressed and negligible in our
 sample.} 
\label{fig:mass_pipi0eta.eps}
\end{figure}

Our branching fraction for $\tau^-\to\pi^-\pi^0\eta\nu_{\tau}$ decay is
consistent with the prediction based on CVC and experimentally measured
$e^+e^-\to\pi^+\pi^-\eta$ cross sections~\cite{CVC}. In addition, the
Monte Carlo code TAUOLA reproduces the observed 
hadronic mass distributions rather well as shown 
in Fig.~\ref{fig:mass_pipi0eta.eps} while more tuning is needed 
for $\tau$'s decay modes involving kaon(s).
The values of the branching fractions obtained for $\tau^-\to
K^-\eta\nu_{\tau}$ 
and $\tau^-\to K^-\pi^0\eta\nu_{\tau}$ decays differ slightly from
the predictions by Li~\cite{Li}.

Further studies of final state dynamics and resonance
formation in the $\tau^- \to K^-\pi^0\eta\nu_{\tau}$ and
$\pi^-\pi^0\eta\nu_{\tau}$ decays, which may be important for understanding the WZW anomaly are
in progress.

\end{document}